\documentclass[letterpaper]{article}
\usepackage{aaai}
\usepackage{times}
\usepackage{helvet}
\usepackage{courier}
\usepackage{multirow}
\usepackage{array}
\usepackage{url}
\usepackage{comment}
\usepackage{subfigure}
\usepackage{graphics}
\usepackage{epsfig}
\usepackage{amssymb,amsmath,amsfonts}

\begin{document}

\title{Predicting Successful Memes using Network and Community Structure}

\author{
Lilian Weng \and Filippo Menczer \and Yong-Yeol Ahn\\
Center for Complex Networks and Systems Research\\
School of Informatics and Computing\\
Indiana University, Bloomington, USA\\
}

\maketitle

\begin{abstract}

We investigate the predictability of successful memes using their early spreading patterns in the underlying social networks. We propose and analyze a comprehensive set of features and develop an accurate model to predict future popularity of a meme given its early spreading patterns. Our paper provides the first comprehensive comparison of existing predictive frameworks. We categorize our features into three groups: influence of early adopters, community concentration, and characteristics of adoption time series. We find that features based on community structure are the most powerful predictors of future success. We also find that early popularity of a meme is not a good predictor of its future popularity, contrary to common belief. Our methods outperform other approaches, particularly in the task of detecting very popular or unpopular memes. 

\end{abstract}


\section{Introduction}

A \emph{meme} is a piece of information that replicates among people~\cite{meme}. Memes bear  similarities to infectious diseases, as both travel through social ties from one person to another~\cite{daley_epidemics_1964,goffman1964nature}. The wide adoption of online social networks not only makes \emph{Internet memes} possible, but also provides us with valuable data on the spreading of memes and user behavior~\cite{Vespignani2009,Lazer2009}.
Although numerous memes are created everyday, only a few go \emph{viral}, prompting a question that has attracted attention across disciplines including marketing, advertisement, and social media analytics, as well as machine learning and network science: \emph{can we predict successful memes at their early stage?}

What makes a meme viral? First, a meme may become viral simply because the meme appeals to many~\cite{Berger:2009viral,cataldi2010}.  At the same time, given the competition between memes and social influence, innate appeal alone may not be able to paint the whole picture~\cite{Salganik2006music,Kitsak2010kcore,Mei2012tags}. The success of a meme also depends on timing, network structure, randomness, and many other factors~\cite{Centola2010SpreadExp,Weng:2012scirep,pinto2013using}.

We identify two major approaches to meme virality prediction: time series analysis and feature-based classification. 
Time series analyses focus on the patterns of early popularity fluctuation of a meme, assuming that the patterns of a meme's growth and decay tell us whether it will go viral in the future~\cite{Jamali2009digging,asur_trends_2011,Yang2011temporal}. 
Classification approaches commonly aim to discover distinguishing features of successful memes by applying supervised machine learning techinques with labled datasets. 
A variety of features have been proposed and tested to differentiate viral memes from others; examples include comments, votes, and user-defined groups~\cite{Lerman2010www,Jamali2009digging,Suh2010,Hong2011popu,Mei2012tags}. However, most studies have paid little attention to the role of the underlying network structure~\cite{romero2011popu,ma2013predicting} even though it is natural to expect network topology to affect information diffusion, as memes spread through social ties.

Here we demonstrate that features based on network structure, particularly \emph{communities}---defined as densely connected clusters of people~\cite{fortunato2010community}---provide crucial insights into virality. We offer the first comprehensive comparative analysis of three categories of features: 
The first category includes features that capture the \emph{audience size}. As many studies on social influence have assumed, the neighbors of an individual in the network can be considered as their potential audience~\cite{Kitsak2010kcore,cha2010measuring,Suh2010,bakshy_everyones_2011}.  For example, one of the common beliefs is that star users with lots of followers are more influential than others with fewer followers.
Second, we examine the predictive power of \emph{community structure}, as it was shown that the spreading pattern of a meme across communities reveals the general appeal of the meme~\cite{weng2013viral}.
Finally, we take into account the \emph{speed of growth} in early meme adoption.

By comparing with multiple representative prediction models, we show that our model can accurately predict the popularity of memes (to an order of magnitude) \emph{two months in advance}, with knowledge of only a small number of early tweets. Our model outperforms random guessing, majority guessing, and three regression models that use early popularity or expected influence of early adopters.

\section{Background}

The virality of memes has been examined from various perspectives, including their innate attractiveness and the role of influentials along with their adoption patterns.

The innate appeal of a meme is commonly believed to contribute to its virality. \citeauthor{Berger:2009viral}~\shortcite{Berger:2009viral} studied emotion hidden in the content of news articles and found that, for instance, articles that evoke arousal are more viral. \citeauthor{Guerini2011text}~\shortcite{Guerini2011text} characterized various aspects that indicate the virality of text-based content, based on the assumption that virality is an intrinsic trait of content. 
\citeauthor{tsur2012wsdm}~\shortcite{tsur2012wsdm} analyzed a rich set of content-based features extracted from hashtags, such as the number of words contained, spelling, lexical items, location in tweets, emotional and cognitive dimensions, in order to predict future popularity of the hashtags.
Yet, randomized experiments on music choices and social news filtering suggested that innate quality may play only a minor role in
determining future popularity due to the strong effect of social influence~\cite{Salganik2006music,muchnik2013}.

User behaviors and characteristics are other important aspects. Limited individual attention causes competition
among memes, inducing strong heterogeneity in meme popularity and 
longevity~\cite{Weng:2012scirep}. Each user has different interests affecting 
adoption preference and meme popularity~\cite{Mei2012tags}.
Many methods for quantifying user influence and identifying influential users---\emph{influentials}---have been proposed. User influence is often quantified in terms of high degree or retweetability~\cite{cha2010measuring,Suh2010}, topical
similarity~\cite{tang2009social,weng2010twitterrank}, information forwarding
activity~\cite{romero2011influence,Suh2010}, or size of cascades~\cite{Kitsak2010kcore,bakshy_everyones_2011}. Here we evaluate our model against a baseline built upon social influence.

The structure of the underlying networks has been shown to have a significant impact on the spreading process in general~\cite{daley_epidemics_1964,goffman1964nature,vesp2008book,vespignani2001SISonSF} and vice versa~\cite{weng2013shortcut}. The existence of hubs---nodes with lots of neighbors---is known to affect the persistence of infections, the distribution of cascade sizes, and the vulnerability of the system~\cite{vespignani2001SISonSF,Watts2002}. Other important network structures present in most real networks are dense subgraphs called \emph{communities}~\cite{newman2006modularity,Rosvall2008InfoMap,Ahn2010LinkCommunity,fortunato2010community}. Communities are believed to constrain information flow or the spreading of diseases~\cite{weak_ties_1973,JP2007pnas,Rosvall2008InfoMap,colbaugh2012early,weng2013viral}.

The spread of memes is often considered as \emph{social contagion}, commonly defined as the spread of information or behavior on social networks where an individual serves as the stimulus for the imitative actions of another~\cite{contagion1985,goffman1964nature,daley_epidemics_1964}. However, studies have shown that information contagion may spread differently from diseases, as multiple exposures can significantly increase the chances of adoption~\cite{threshold_model,Centola2010SpreadExp,romero_differences_2011}.
The speed and ease of meme transmission is affected by characteristics of social ties.  Strong and homophilous ties are often seen as more effective than weak ties for spreading messages~\cite{Brown:1987kx}, while weak ties are expected to transmit novel information~\cite{weak_ties_1973}.  In viral marketing and consumer studies, researchers actively apply network approaches to analyze and model local and global structural patterns of social networks~\cite{Leskovec2007viral,Mason2008propagation,aral2011viral}.

One of the common approaches to detect viral memes is \emph{time series analysis}, which examines temporal patterns such as growth, bursts, and decay~\cite{wu_novelty_2007,romero_differences_2011,asur_trends_2011}. A common finding is that temporal patterns of memes can be well summarized into a few categories, and they have predictive power to spot trendy or bursty memes~\cite{Yang2011temporal,Cattuto2012WWW}. Classification of temporal patterns can be seen as an extended application of trajectory clustering~\cite{Gaffney1999KDD,Lee2007TrajClus}.  Existing virality prediction algorithms try to forecast time series based on past values~\cite{MacNames1998,Lenser2005ICRA,Kaltenbrunner2007Slashdot}.  
Some event detection methods group memes together to form topics and use temporal activity to detect trending topics~\cite{Becker2011icwsm,cataldi2010}.

In another approach, the prediction problem is treated as a
\emph{classification} task. Multiple studies have claimed that the early popularity
of online content is strongly correlated with its future
popularity~\cite{Jamali2009digging,Szabo2010pred,Lerman2010www}. 
\citeauthor{Szabo2010pred}~\shortcite{Szabo2010pred} proposed a model that predicts future popularity based on early popularity. \citeauthor{Jamali2009digging}~\shortcite{Jamali2009digging} used daily user
activities, user interest peak, and comment information attached to each Digg
story to estimate future usage. Design elements of
a website are shown to be informative as well; \citeauthor{Lerman2010www}~\shortcite{Lerman2010www} found that 
incorporating design features of the website can improve
the outcomes of their stochastic prediction model.
The numbers of URLs and hashtags in a tweet are suggested to be strongly correlated with its retweetability, while the number of followers, followees, and the account age have a weak effect~\cite{Suh2010}. \citeauthor{Mei2012tags}~\shortcite{Mei2012tags} quantified how a user selects content tags using individual interests, relevance, and behavior of neighbors; however, the features are proposed for predicting whether a single user will adopt a given hashtag, not applicable for foretelling the future popularity of hashtags.
Some other notable features include content properties such as terms, language, semantics, and category~\cite{tsur2012wsdm}, user influence~\cite{bakshy_everyones_2011,Salganik2006music}, source authority~\cite{Bandari2012pred}, and the graph topology of early adopters~\cite{romero2011popu,ma2013predicting}.
In a recent paper, \citeauthor{Cheng14CanCascadesBePredicted}~\shortcite{Cheng14CanCascadesBePredicted} formulated social virality prediction as a sequence of binary classification problems, while a cascade is tracked over time. In spite of the different problem formulation, our results seem to be consistent with their finding that initially, breadth is a strong indicator of larger cascades.

\section{Dataset}

\begin{table}
\caption{Basic statistics of the reciprocal follower network in the study. Node coverage measures the proportion of nodes belonging to communities that have at least three nodes.}
	\centering
	\begin{tabular}{ p{2cm} l >{\hfill}p{2cm}}
	\hline
	& \#Nodes  &  400,020 \\
	& \#Edges  & 10,012,989 \\
	& Clustering coefficient &  0.2093 \\
	\hline
	\multirow{2}{*}{InfoMap} & \#Communities &  6,569 \\
	& Node coverage & 99.08\% \\
	\multirow{2}{*}{LinkClustering} & \#Communities & 193,805\\
	& Node coverage & 43.30\% \\
	\hline
	\end{tabular}
\label{table:dataset}
\end{table}

Here we harnessed a dataset from Twitter, one of the most popular micro-blogging
platforms, where users post short posts called \emph{tweets}. Twitter provides
a great opportunity to study the spread of memes because (i)~it is one of the main
platforms where internet memes are generated and shared, and (ii)~it
supplies network structure, content of messages, spreading events, and
ways to define memes concretely. 
Between a pair of users $(u,v)$, we consider three main types of interactions:
(i) $u$ can \emph{follow} $v$ to subscribe to $v$'s activities (tweets, retweets, etc.); (ii) $u$ can
\emph{retweet} $v$'s messages to re-broadcast it to $u$'s followers, commonly
noted as ``RT'' for short; (iii) $u$ can \emph{mention} $v$'s screen name in tweets by
using the ``@'' symbol (e.g. `@yy').  Users can also explicitly attach
indexable topic identifiers to a tweet by using \emph{hashtags}, topical terms
with the ``\#'' symbol as a prefix (e.g. `\#news').

We consider each hashtag as a meme, as we can concretely identify and track
hashtags, and as properties of hashtags accord with the definition of
meme~\cite{meme}; most hashtags are unique phrases that spread by imitation.
Moreover, they mutate, compete, and survive; Twitter users quickly reach
consensus on representative hashtags for certain topics.  For instance,
\texttt{\#ows} quickly became \emph{the} hashtag of the Occupy Wall Street
movement---outcompeting similar ones---among hundreds of thousands of people
who participated in public discourse around the movement~\cite{conover2013ows}.

By using the Twitter Streaming API and the `GET followers' method of the Twitter REST API, we collected tweets during March and April of 2012 and reconstructed a relevant portion of the follower network. We only kept reciprocal follow links, as bi-directional communication reflects more stable and reliable social connections. Although we expect that incorporating the direction of connections may improve our results even more, we stick with bi-directional links for the sake of simplicity and generality. Non-English users were filtered out to avoid any artifact from the large-scale segregation between language groups.

We identified communities on the resulting network by using two algorithms: InfoMap~\cite{Rosvall2008InfoMap} and LinkClustering~\cite{Ahn2010LinkCommunity}, to demonstrate the robustness of our experiments against specific choices of community detection methods. 
We have chosen these methods primarily because of their performance, and partly because they are based on contrasting principles, in order to confirm the robustness of the results; InfoMap detects disjoint communities while LinkClustering identifies overlapping communities. 
In our analysis we ignore communities with fewer than three nodes.
Basic statistics of the network and communities are displayed in Table~\ref{table:dataset}.

\section{Definitions}

Let us first define key concepts and mathematical notations to facilitate the subsequent discussion.

\noindent \textbf{Definition~1. Meme and meme popularity:} 
We consider each hashtag $h$ as a meme. 
$T(h)$ is a set of all tweets that contain $h$ and $T_n(h)$ is a set of the earliest $n$ tweets that contain $h$. 
Thus $T_n(h) \subseteq T(h)$ and $n = |T_n(h)| \leq |T(h)|$.  
Similar definitions can be made for adopters. 
$A(h)$ is a set of all adopters who tweeted about $h$ and $A_n(h) \subseteq A(h)$ is a set of early adopters who tweeted at least one of the first $n$ tweets. 
The popularity of meme $h$ is quantified by the number of tweets, $|T(h)|$, or adopters, $|A(h)|$. 

\noindent \textbf{Definition~2. Network surface:} 
The neighbors of a given set of users $U$ (not counting $U$) are deemed to be $U$'s \emph{surface} $S(U)$. 
The definition of the surface can be extended recursively to the $k$-th surface, which contains users within $k$ steps from any user in the target set $U$, $S^k(U) = S(S^{k-1}(U)) \cup S^{k-1}(U)$, and $S^1(U) = S(U)$. 

\noindent \textbf{Definition~3. Adopter sequences and time series:} 
For a given meme $h$, we consider the sequence of meme adopters, $\langle a_1^h, a_2^h, \dots, a_{|T(h)|}^h \rangle$, where $a_i^h \in A(h)$ is the creator of the $i$-th tweet with $h$. 
A user may appear multiple times in the sequence if the user tweets about $h$ more than once. 
Similarly we build the tweet time series $\langle t_1^h, t_2^h, \dots, $ $t_{|T(h)|}^h \rangle$ where $t_i^h$ marks the timestamp (in second) of the $i$-th tweet containing $h$.
The set of tweets within time $\tau$ is labeled $T^{\tau}(h)$ where $\tau$ is a time duration measured starting from the first tweet.

\noindent \textbf{Definition~4. Community:} 
A community $c \in \mathcal{C}$ is a subset of nodes (users) in the network. 
$T(h|c)$ and $A(h|c)$ are tweets and adopters of a meme $h$ in community $c$, respectively. 
We define $T_n(h|c)$ and $A_n(h|c)$, that consider only early tweets, in a similar fashion. 
$C(h)$ denotes the \emph{infected communities} of $h$, which generate at least one tweet containing $h$; $C(h) = \{c \; | \; c \in \mathcal{C}, |T(h|c)| \geq 1 \}$.  
Similarly, the infected communities with early tweets are $C_n(h) = \{ c \; | \; c \in C(h), |T_n(h|c)| \geq 1 \}$. 

\noindent \textbf{Definition~5. Interactions:} 
$I(h)$ is the number of user interactions regarding $h$. Two types of user interactions are considered: retweets ($\mathrm{RT}$), by which a user retweets a message containing $h$ from another user; and mentions ($\mathrm{@}$), by which a user mentions another in a tweet containing $h$.
We consider interactions within communities, $I^{\circlearrowright}(h)$, and between communities, $I^{\curvearrowright}(h)$, respectively, where $I(h) = I^{\circlearrowright}(h) + I^{\curvearrowright}(h)$.

\section{Characterizing Viral Memes}

In this section we identify signatures of viral memes at their early stages in terms of three characteristics: \emph{network topology}, \emph{community diversity}, and \emph{growth rate}.  
We demonstrate that the information on early adopters, particularly in the context of social network structure, is powerful enough to identify \emph{young} viral memes. Let us present the rationales for the prediction features used in the model before introducing the detailed definition of each feature in the next section.

\subsection{Network Topology}

\begin{figure}[t]
\centerline{\includegraphics[width=0.85\columnwidth]{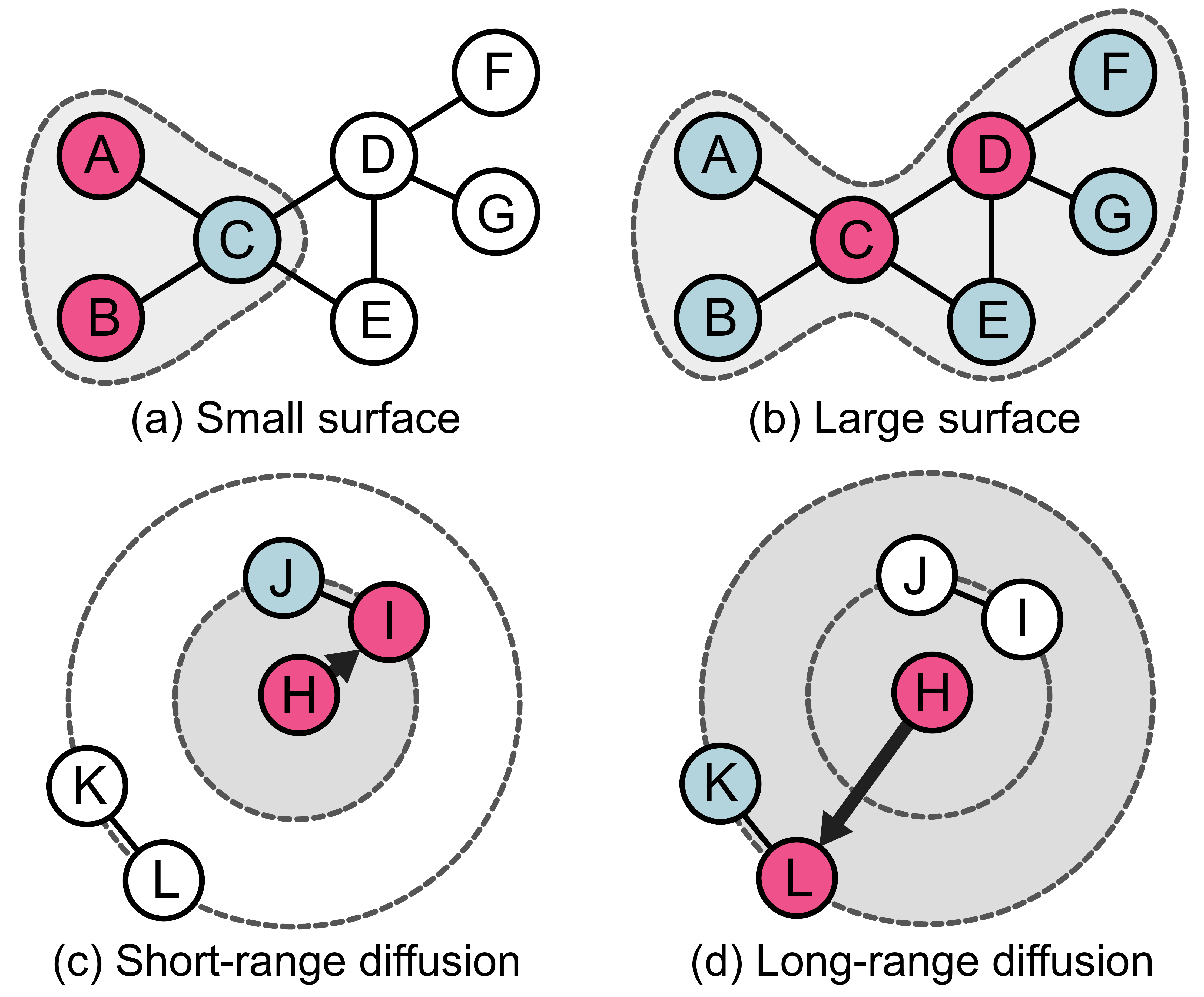}}
\caption{Network surfaces. Red nodes represent the early
adopters of a meme; blue ones are neighbors of early adopters;
the grey color marks the surface area. 
(a) When $A$ and $B$ adopt a meme, the corresponding
network surface is small.
(b) The adoption by $C$ and $D$ creates a large network surface. 
(c) Adopter $H$ spreads the meme to node $I$ (a nearby node), and the potential 
adopters do not change much. 
(d) When $H$ spread the meme to node $L$ (a node farther away),
the set of potential meme adopters grows a lot.} 
\label{fig:net_viz}
\end{figure}

The position of an adopter in the network determines the size of the potential audience~\cite{Kitsak2010kcore}. 
The network surface of a given set of adopters, $S$, captures the
number of neighbors who are directly exposed. As illustrated in
Fig.~\ref{fig:net_viz}(a-b), the network surface varies greatly depending on
the degrees and positions of the adopters. 
We also estimate the growth of potential audience in time by examining the distance between consecutive adopters in the network. Note that new adopters are not necessarily connected to existing adopters because a meme can be injected into multiple nodes of the network, and because our collection is based on a sample of the entire public stream. The farther the jump between two consecutive adopters, the more potential spreaders the meme may have (Fig.~\ref{fig:net_viz}(c-d)).  

\subsection{Community Diversity}

To explain our community diversity features, let us first examine the characteristics of social contagion. Unlike epidemic diseases, social contagions are known to possess two distinctive characteristics: 

\begin{description}

\item[Social reinforcement.] Until a certain point, each additional exposure drastically increases the probability of adoption~\cite{Centola2007weakTies,bakshy2009adoption,romero_differences_2011,Centola2010SpreadExp}.

\item[Homophily.] Social relationships are more likely to be formed between people who share characteristics, captured in the sayings ``birds of a feather flock together'' and ``similarity breeds connection''~\cite{McPherson2001homophily,Centola2011homophily}. Therefore, we expect to see that connected people have similar characteristics, such as interests, languages, or culture, increasing the chances of adopting similar memes.  

\end{description}

\begin{figure}[t]
\centering
\includegraphics[width=\columnwidth]{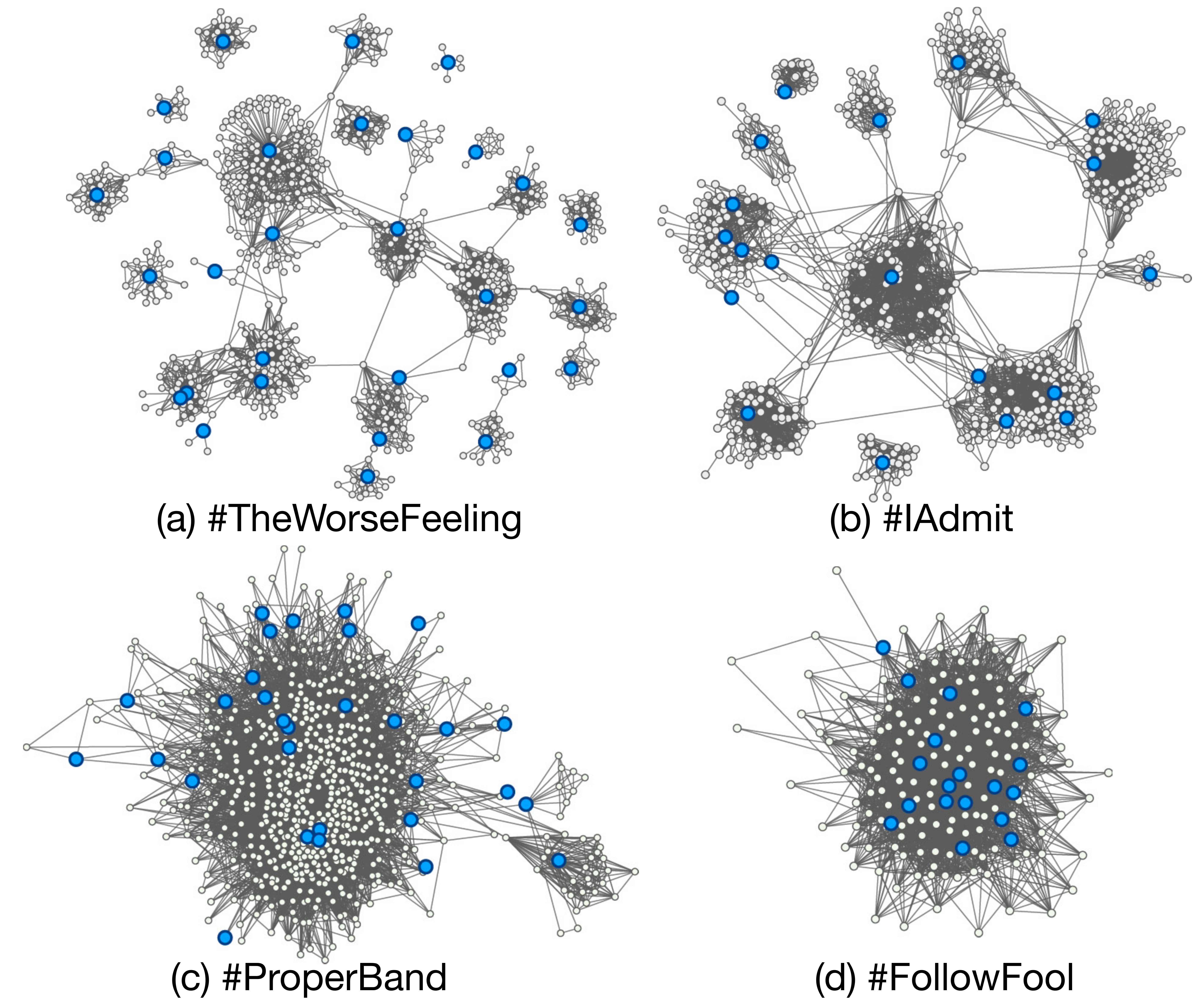}
\caption{Visualizations of diffusion patterns of  viral (a,b)
and non-viral (c,d) memes. The early adopters among the first 30
tweets and their neighbors in the same communities are shown. Each
node represents a user and each link indicates the reciprocal follow
relationship between two users. Adopters are colored in blue.
} \label{fig:comm_viz}

\end{figure}

Community structure has been shown to help quantify the strength of these effects~\cite{colbaugh2012early,weng2013viral}. 
First, dense connectivity inside a community increases the chances of multiple exposures, thus enhancing the contagion that is sensitive to social reinforcement. Second, groups with similar tastes naturally establish more edges among them, forming communities. Therefore members of the same community are more likely to share similar interests. We thus expect that, if these two effects are strong, communities will facilitate the internal circulation of memes while preventing diffusion across communities, causing strong concentration or low \emph{community diversity}.
Our previous study showed that unpopular memes tend to be concentrated in a small number of communities while few viral memes have high community diversity, spreading widely across communities like epidemic outbreaks~\cite{weng2013viral}. We expect that features that quantify the community diversity should help predict future meme virality. As an illustration, we visualize the early diffusion patterns of a few memes based on the first 30 tweets in Fig.~\ref{fig:comm_viz}. Viral hashtags such as \texttt{\#TheWorseFeeling} and \texttt{\#IAdmit}, exhibit more community diversity than non-viral me\-mes, e.g. \texttt{\#ProperBand} and \texttt{\#FollowFool}.

\subsection{Meme Growth Rate}

\begin{figure}
\centering
\includegraphics[width=\columnwidth]{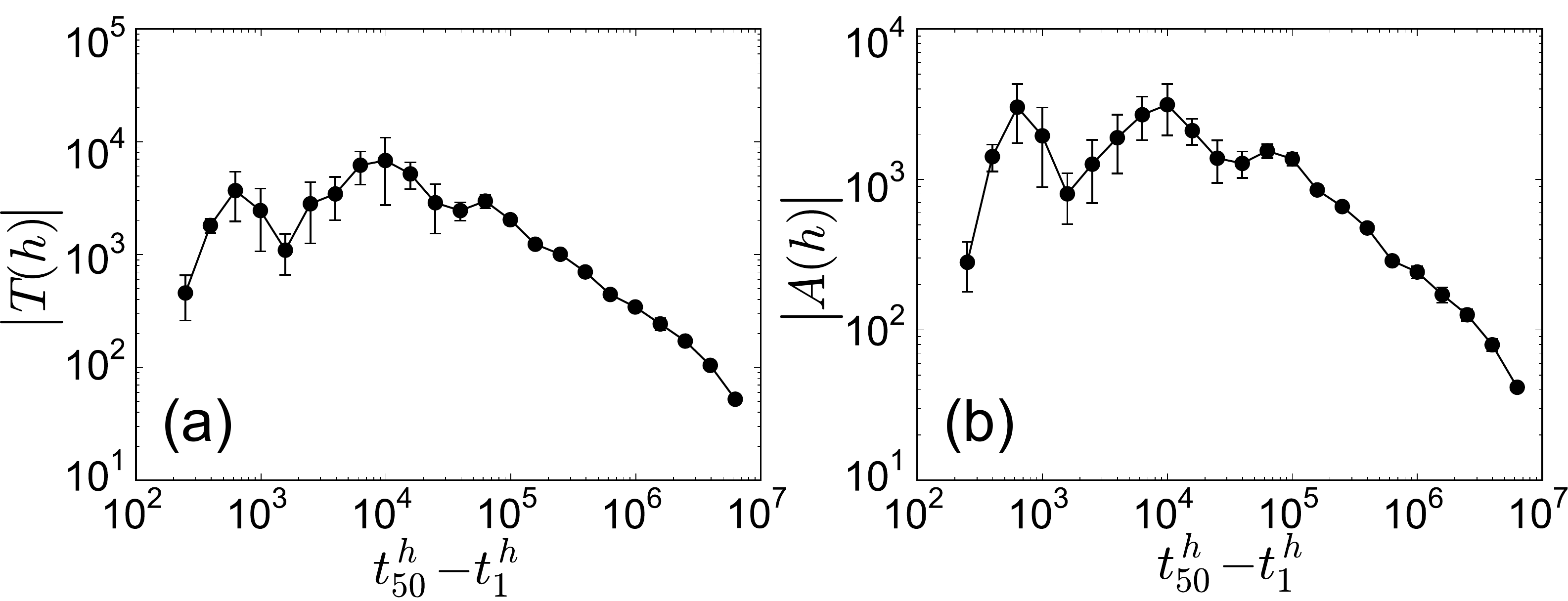}
\caption{The relationship between the meme popularity measured in the number of tweets, $|T(h)|$, and adopters, $|A(h)|$, and the early spreading time with $n=50$, $t_{50}^h - t_1^h$ seconds.} 
\label{fig:ts_viz}
\end{figure}

Viral memes are expected to spread more quickly than others~\cite{Szabo2010pred}. 
To incorporate this intuition, we define the time difference between the first and the $n$-th tweet in the time series of a meme $h$ as the \emph{early spreading time}, $t_n^h - t_1^h$. It gauges the initial \emph{growth rate} of $h$.  
Figure~\ref{fig:ts_viz} displays a correlation between the growth rate and meme popularity. 
Although we observe fluctuations when the early spreading time $t_{50}^h - t_1^h$ is small, meme popularity significantly decreases when the early spreading is slow. 

\section{Prediction Features}\label{sec:features}

Based on our preliminary analyses above, we design features for our prediction model. 
Network features describe the size of potential audience based on the positions of early adopters in the network.  
Community features measure the community diversity at the early stage.  
Growth-rate features quantify the initial momentum.  
We have 13 features in total, marked as $f$.1--13. All the features are computed based on the first $n$ tweets for each hashtag, where the parameter $n$ is a small number compared with the final number of tweets generated by viral hashtags.  


\subsection{Basic Network Features}

Here we use the connectivity of users.

\noindent $f$.1. \textbf{Number of early adopters}, $|A_n(h)|$.  
The number of early adopters is one of the most basic and simple features. 
$A_n(h)$ is the set of distinct adopters in the earliest $n$ tweets of a meme $h$. 
A small $|A_n(h)|$ would indicate that a small number of users generated most tweets and the hashtag is failing to spread.

\noindent $f$.2. \textbf{Size of first surface}, $| S(A_n(h)) |$. The first surface contains all the uninfected neighbors of early adopters of $h$. It is the set of most immediate adopter candidates~\cite{ma2013predicting}. 

\noindent $f$.3. \textbf{Size of second surface}, $| S^2(A_n(h))  |$. The second surface includes uninfected users in the second surface of early adopters, characterizing the number of potential adopters within two steps. 

\subsection{Distance Features}

Here we use the position of adopters in the network.

\noindent $f$.4. \textbf{Average step distance}, $\overline{d_n(h)}$.  With the
adopter sequence for the first $n$ tweets of $h$, $\langle
a_1^h, a_2^h, \dots, a_n^h \rangle$, we measure the 
shortest network path length between consecutive users  
and call it \emph{step distance} $d(a_i^h, a_{i+1}^h)$, where $1 \leq
i \leq n-1$. 
We examine the average distance between consecutive adopters of $h$ in time:
\begin{equation*}
\overline{d_n(h)} = \frac{1}{n-1} \sum_{i=1}^{n-1} d(a_i^h, a_{i+1}^h).
\end{equation*}

\noindent $f$.5. \textbf{CV of step distances}, $C_v(d_n(h))$.  The coefficient of
variation ($C_v$) of a variable is the ratio of its standard deviation to the
mean. We use it to measure the relative variability in step distance:
\begin{equation*}
C_v(d_n(h)) = \frac{1}{\overline{d_n(h)}} \sqrt{\frac{\sum_{i=1}^{n-1}(d(a_i^h,
a_{i+1}^h) - \overline{d_n(h)})^2}{n-2}}.
\end{equation*}

\noindent $f$.6. \textbf{Diameter}, $D_n(h)$.  The diameter is the maximum
distance between any two adopters of $h$ within the first $n$ tweets.  
It is a measure of audience coverage in the network:
\begin{equation*}
D_n(h) = \max_{1 \leq i \neq j \leq n-1} d(a_i^h, a_j^h).
\end{equation*}

\subsection{Community Features}

Community-based features are designed on the basis of our previous study, showing that viral memes exhibit high community diversity~\cite{colbaugh2012early,weng2013viral}. That study includes a scrupulous assessment of these community-based features and a comparison with results produced by several synthetic diffusion models.
The features are computed at prediction time, based on the predefined communities; the community detection algorithm is executed once on the network built upon the historical data, as the network structure does not evolve much within a short time period.

\noindent $f$.7. \textbf{Number of infected communities}, $|C_n(h)|$.  It is the number of communities with at least one adopter of $h$ among first $n$ tweets.





\noindent $f$.8--9. \textbf{Usage and adopter entropy}, $H_n^T(h)$ and $H_n^A(h)$.
The measurement of entropy describes how  tweets or adopters of a
given meme are scattered or concentrated across communities. Large entropy indicates high
diversity and low concentration:
\begin{eqnarray*}
H_n^T(h) & = & -\sum_{c \in C(h)} \frac{|T_n(h|c)|}{n} \log
\frac{|T_n(h|c)|}{n} \\
H_n^A(h) & = & -\sum_{c \in C(h)} \frac{|A_n(h|c)|}{|A_n(h)|} \log
\frac{|A_n(h|c)|}{|A_n(h)|}.
\end{eqnarray*}

\noindent $f$.10--11. \textbf{Fraction of intra-community user interaction},
$I_n^{\circlearrowright}(h)/I_n(h)$.  The likelihood of a user adopting information
from members of the same community increases with the strength of the community
trapping effect. We expect to observe weaker community trapping and higher
community diversity in early adopters of viral memes.  Here we quantify this by
measuring the fraction of intra-community user interaction. The interactions
can be retweets or mentions:
\begin{eqnarray*}
\frac{I^{\circlearrowright\mathrm{RT}}_n(h)}{I^\mathrm{RT}_n(h)} \quad , \quad
\frac{I^{\circlearrowright\mathrm{@}}_n(h)}{I^\mathrm{@}_n(h)}.
\end{eqnarray*}
A high fraction of intra-community interaction suggests a limited group of potential adopters in the future. 

\subsection{Growth Rate Features}

Given the time series of the first $n$ tweets of a meme $h$, $\langle t_1^h,
t_2^h, \dots, t_n^h \rangle$, we can measure \emph{step time duration}---the
time difference between consecutive tweets, $t_{i+1}^h - t_i^h$. The mean 
and fluctuations of the sequence of time durations are implemented as two prediction
features.

\noindent $f$.12. \textbf{Average step time duration}, $\overline{\Delta t_n(h)}$:
\begin{equation*}
\overline{\Delta t_n(h)} = \frac{\sum_{i=1}^{n-1} t_{i+1}^h - t_i^h}{n-1}  =
\frac{t^h_n - t^h_1}{n-1}.
\end{equation*} 

\noindent $f$.13. \textbf{CV of step time durations}, $C_v(\Delta t_n(h))$:
\begin{equation*}
C_v(\Delta t_n(h)) = \frac{1}{\overline{\Delta
t_n(h)}}\sqrt{\frac{\sum_{i=1}^{n-1} (t_{i+1}^h - t_i^h - \overline{\Delta
t_n(h)})^2}{n-2}}.
\end{equation*}

\section{Experiments}
\label{sec:exp}

In this section we predict the magnitude of a meme's future popularity using the features introduced above, calculated on the basis of early observation, and compare the results with five baselines.

\subsection{Task Definition}

We define the popularity (virality) of a meme $h$ as the number of tweets $|T(h)|$ or adopters $|A(h)|$. We use both definitions, as they highlight different perspectives of a meme: the former characterizes the amount of discussion a meme triggers; the latter tells us about the size of the crowd participating in the discussion.  Large $T(h)$ does not necessarily implies large $A(h)$, because a single user may generate many tweets.
Meme popularity exhibits a broad and skewed distribution, as observed in many previous studies~\cite{Lerman:ICWSM2010,Weng:2012scirep}. We partition all the memes into classes based on the order of magnitude of the total popularity ($\lceil \log_{10} |T| + 0.5 \rceil$ or $\lceil \log_{10} |A| + 0.5 \rceil$).
The prediction task is therefore a \emph{multi-label classification}. Given the information about the early stage of a hashtag, the task is to predict which class it belongs to after about two months, at the end of the observation period of our dataset.

\subsection{Baselines}

We evaluate our prediction results by comparing them with five baseline prediction models: $B_1$ and $B_2$ are trivial baselines; $B_3$, $B_4$, and $B_5$ are regression models that use features such as social influence of adopters, cumulative popularity, and the growth sequence of memes. Note that content-based prediction models, such as the model proposed by \citeauthor{tsur2012wsdm}~\shortcite{tsur2012wsdm}, are not considered as we focus on the prediction problem using only network spreading patterns, without looking into the content.

\begin{enumerate}

\item Random guess ($B_1$): Assuming that we know the exact number of memes in
    each class, $B_1$ randomly assign the class label to each meme with the prior
    probability.

\item Majority guess ($B_2$): Due to the imbalanced distribution of meme
    popularity, simply assigning the dominant class label to every meme yields
    high accuracy. Note that, however, $B_2$ fails to capture the most important
    but not dominant class---the most viral memes.  This simple but `powerful' 
    baseline has been ignored in most existing studies.


\item Social influence model ($B_3$): 
This is built on the common notion that influential users play a key role in the wide adoption of a meme~\cite{Kitsak2010kcore,cha2010measuring,Suh2010,bakshy_everyones_2011}.
We calculate each user's PageRank score~\cite{brin1998anatomy} and number of followers, which approximately captures the importance of the user in the network and the size of potential viewers of his content, respectively. 
 According to the social-influence perspective, if a meme is reposted by more influential people at the early stage, it is more likely to go viral. 
For each given meme, we therefore compute the maximum, mean, median, and coefficient of
variation among PageRank scores of its $n$ early adopters; a feature set is built similarly for the follower count, but on a logarithmic scale. We then apply multivariate linear regression using these eight features as one of the baseline models.

\begin{figure*}
\centering
\includegraphics[width=0.95\textwidth]{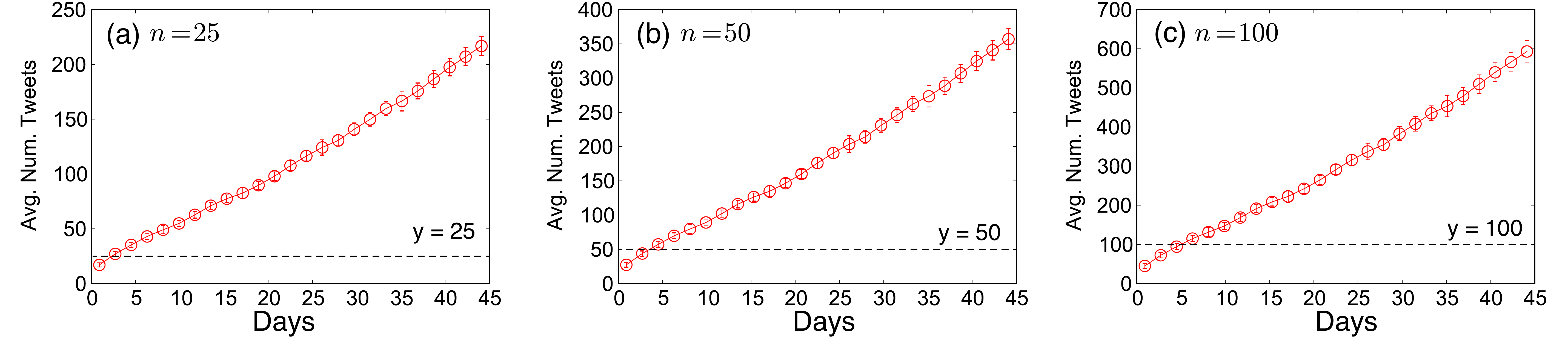}
\caption{The average number of tweets for memes with a given minimum $n$ as a function of time since creation. The dashed lines mark where memes get $n$ tweets. We consider (a) $n=25$, (b) $n=50$, and (c) $n=100$.}
\label{fig:LN_model_support}
\end{figure*}

\item LN model ($B_4$): \citeauthor{Szabo2010pred}~\shortcite{Szabo2010pred} proposed a linear regression (LN) model that uses the logarithm of the early popularity of a meme at time $\tau$, $|T^{\tau}|$, to predict its popularity in the future, $|T|$. 
Given that we use parameter values $n=25, 50, 100$, here we set $\tau=7$ days, as it takes, on average, at most 7 days on average to obtain the numbers of tweets required (see Fig.~\ref{fig:LN_model_support}). 

\item ML model ($B_5$): The multivariate linear (ML) model, built upon \citeauthor{Szabo2010pred}'s linear regression model, was proposed by \citeauthor{pinto2013using}~\shortcite{pinto2013using}. Instead of using the cumulative popularity reached by a meme on a given day, the model takes the popularity measured on each day up to time $\tau$ to form a vector as the predictor for the future popularity. We set $\tau =7$ as in $B_4$.


\end{enumerate}

\subsection{Network-based Prediction Model ($P_n$)} 

Since we focus on identifying the predictive features, 
we choose one of the most widely adopted methods---the random forest algorithm---that has been shown to be robust and reliable~\cite{Breiman:2001fk}. We construct 300 decision trees, each with 5 random features from those introduced earlier.  
Our prediction model $P_n$ uses the features computed with the first $n$ tweets of each meme. Note that hashtags with fewer than $n$ tweets are not considered in the calculation. We experiment with multiple values of $n$; the corresponding number of memes in each class is listed in Table~\ref{table:class_stats}.

To ensure that we examine only \emph{new} memes, we only include the hashtags that were used during the first two weeks of Mar 2012 and appeared in less than $X$ tweets during the previous month (Feb 2012); we set $X=20$ for our method as well as the baselines. Our previous study reported that at least the community-based prediction is insensitive to the choice of $X$~\cite{weng2013viral}.

\begin{table}
\caption{The number of `new' hashtags in each class with different $n$ values. Note that only 48 memes in the dataset reach the order of $10^4$ tweets and only 33 memes reach the order of $10^4$ adopters.}
\centering
\begin{tabular}{c | r  r  r  r  | r  r  r  r  r}
\hline
\multirow{2}{*}{$n$} & \multicolumn{4}{c|}{$\lceil \log |T| + 0.5 \rceil$} & \multicolumn{5}{c}{$\lceil \log |A| + 0.5 \rceil$} \\
 & 1 & 2 & 3 & $\geq$4 & 0 & 1 & 2 & 3 & $\geq$4 \\
 \hline\hline
25 & 2,853 & 6,227 & 224 & 48 & 157 & 5,202 & 3,810 & 149 & 33 \\
50 & - & 2,761 & 224 & 48 & 21 & 723 & 2,106 & 149 & 33\\
100 & - & 676 & 224 & 48 & 4 & 118 & 643 & 149 & 33\\
\hline
\end{tabular}
\label{table:class_stats}
\end{table}

\subsection{Evaluation with $F_1$ Score}

\begin{figure}
\centering
\includegraphics[width=\columnwidth]{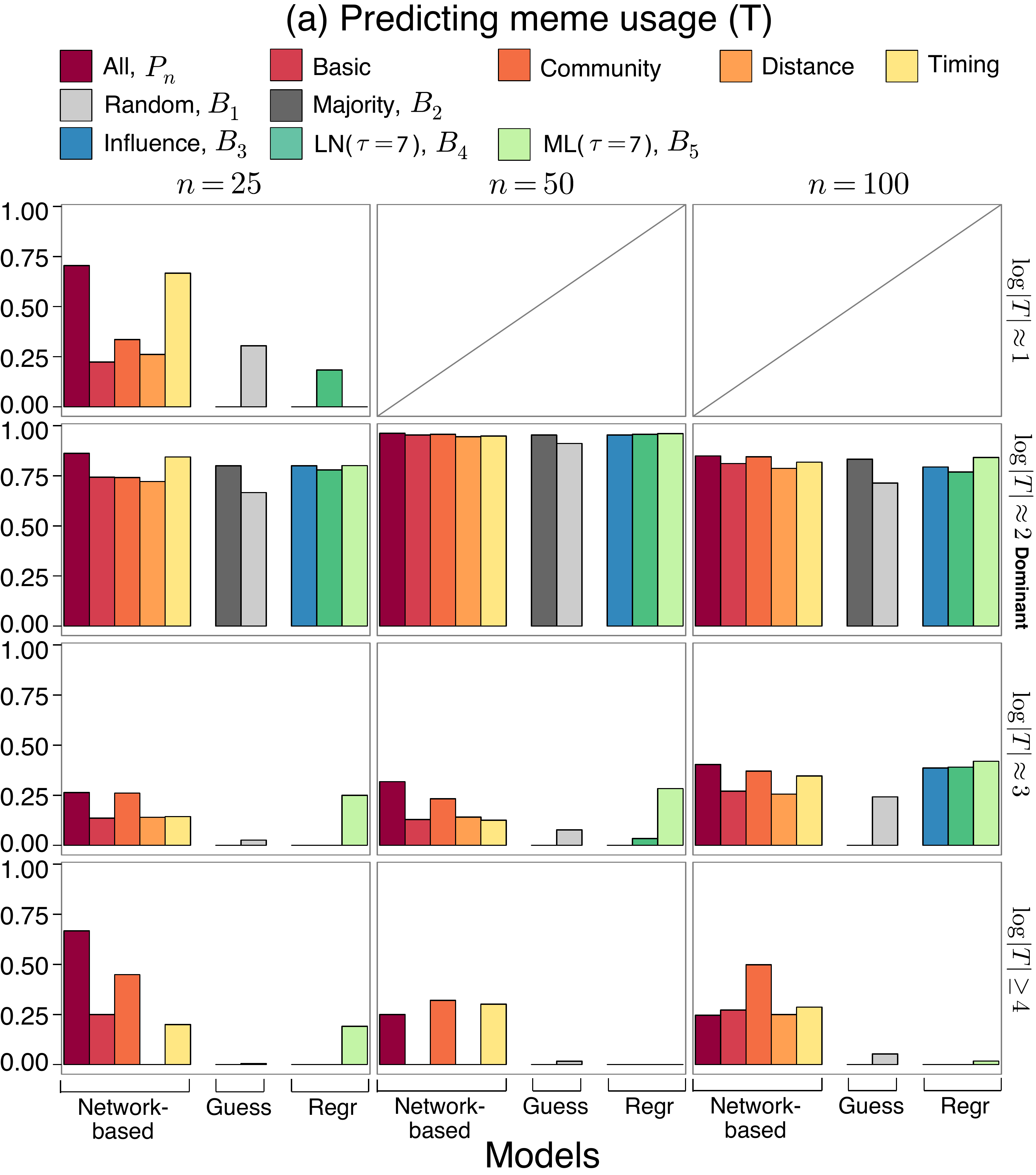}
\includegraphics[width=\columnwidth]{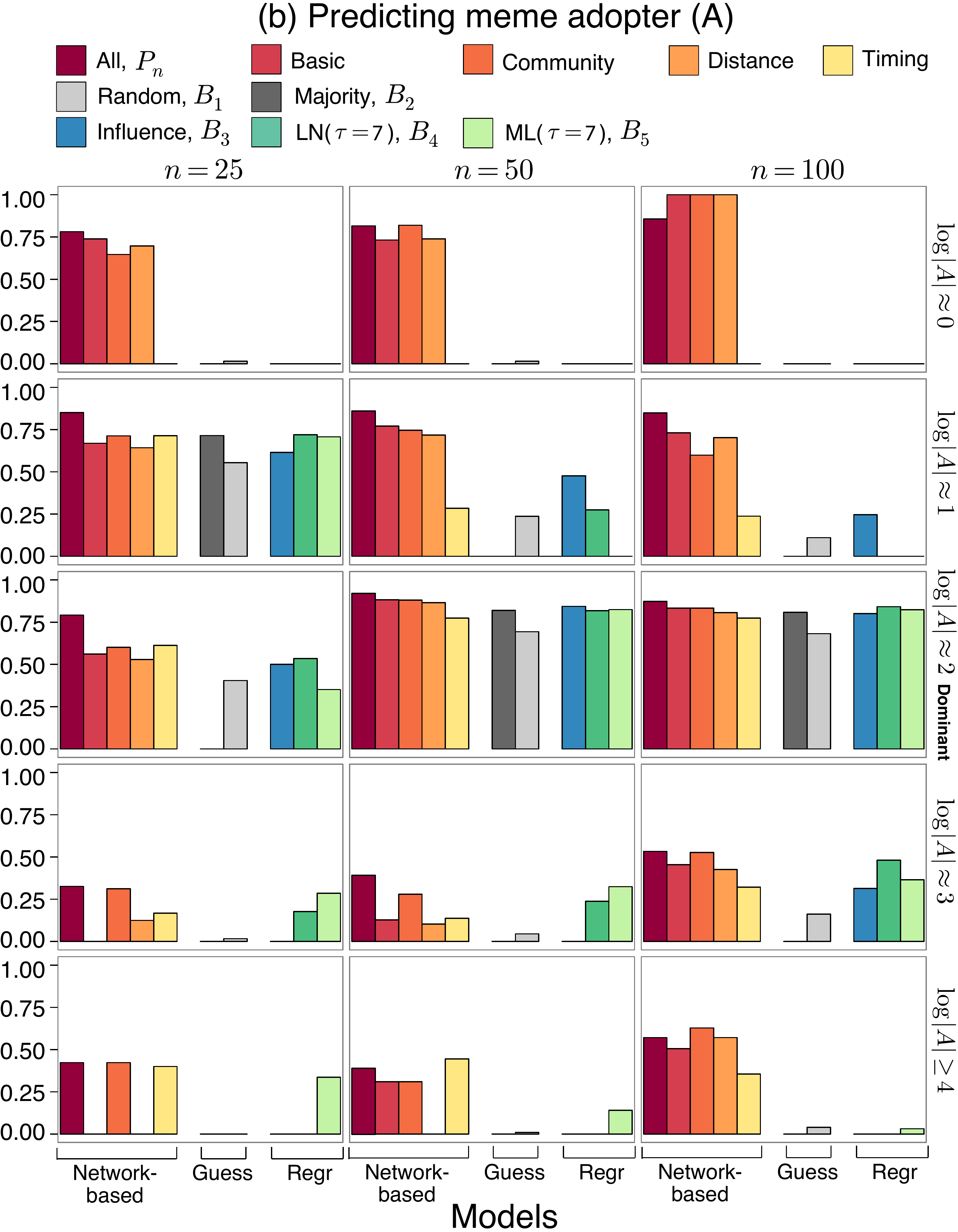}
\vspace{-2em}
\caption{
$F_1$ scores of the prediction models: (a)~$\lceil \log |T| + 0.5 \rceil = 1,2,3,4$ and (b)~$\lceil \log |A| + 0.5 \rceil = 0,1,2,3,4$. The observation window is set to $n=25, 50, 100$ tweets, respectively. Here we only demonstrate the results using the Infomap community detection method; link clustering yields similar results.}
\label{fig:f1score}
\end{figure}


Simply computing the accuracy, the percentage of correctly predicted items among all the items, is not good enough for evaluation in our prediction task, because the classes in our task are imbalanced (see Table~\ref{table:class_stats}). When class sizes are skewed, a high accuracy does not necessarily indicate good performance. Overlooking small classes, as done by the majority guess baseline $B_2$, can yield good accuracy if one or a few dominant classes are over-represented in the dataset.

Instead, we measure both precision and recall for each class to demonstrate the model performance for predicting viral and non-viral memes separately. Precision quantifies how many predicted items for the target class are correct in the empirical data; recall measures how many actual items in the target class are captured by the model.
Precision and recall are combined by the harmonic mean $F_1 = 2 \cdot \text{precision} \cdot \text{recall} / (\text{precision} + \text{recall})$,  between 0 (worst) and 1 (best). $F_1$ scores of different models for predicting the future usage or adopter popularity are displayed in Fig.~\ref{fig:f1score}.
For both $P_n$ and all baselines, we employ 10-fold cross validation. 
To quantify and compare how each set of features in $P_n$ performs, we also run the models with only basic features ($f.1$-$3$), distance features ($f.4$-$6$), community-based features ($f.7$-$11$), and timing features ($f.12$-$13$).

\subsection{Results}

\begin{figure}
\centering
\includegraphics[width=\columnwidth]{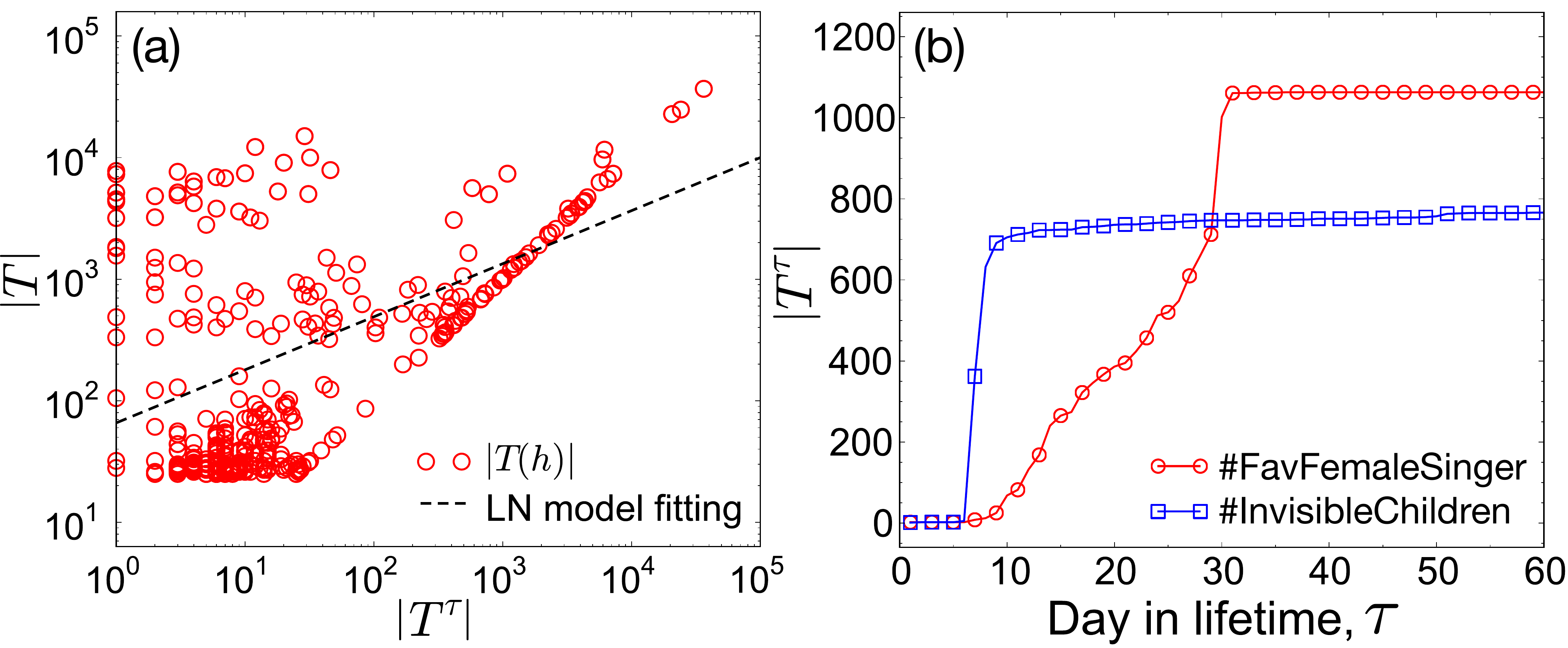}
\caption{(a) Scatter plot of early popularity $T^{\tau}$ versus $T$ for each meme; the black dashed line is the regression line by the LN model. (b) Cumulative popularity for two hashtags, \texttt{\#FavFemaleSinger} and \texttt{\#InvisibleChildren}.}
\label{fig:LN_model}
\end{figure}

All models, including the two trivial baselines ($B_1$ and $B_2$), achieve good results for dominant classes ($\lceil \log |T| + 0.5 \rceil =2$ or $\lceil \log |A| + 0.5 \rceil = 2$), due to the imbalanced class sizes. Note that $B_2$ can only achieve non-zero $F_1$ score in the dominant class. 
Regression models ($B_3$, $B_4$, and $B_5$) in general have similar performance.
We find that the LN baseline model does not work well for the most viral hashtags, because the popularity at the early stage does not guarantee future popularity, in contrast to the common premise of many studies. The correlation between the early popularity $|T^{\tau}|$ and the final popularity $|T|$, as illustrated in Fig.~\ref{fig:LN_model}(a), is weak. This suggests that many initially unpopular hashtags eventually become popular later (cf. upper left quadrant in Fig.~\ref{fig:LN_model}(a)). 
It should be noted that the LN model was originally designed for predicting the popularity of a single piece of online content, such as a Digg story, a YouTube video, or a single tweet, which tends to have swift growth and decay within a shorter lifetime. By contrast, hashtag usage seems to be affected more by long-term endogenous diffusion processes on the network. For instance, in Fig.~\ref{fig:LN_model}(b), the hashtag \texttt{\#FavFemaleSinger} had fewer tweets than \texttt{\#InvisibleChildren} during the first 2 weeks, but it continued to grow and eventually became more popular than \texttt{\#InvisibleChildren}, while \texttt{\#InvisibleChildren} obtained new tweets slowly after the early burst. The LN model may work better for foretelling the future popularity (number of retweets) of a single tweet, but not for hashtags.
The ML baseline ($B_5$) captures more viral memes compared to the other two regression models. The richer description of early growth patterns contained in a meme's daily usage vector yields improved  prediction quality.

Our network-based approach outperforms the five baselines in most cases, especially for the most viral hashtags ($\lceil \log |T| + 0.5 \rceil \geq 4$ or $\lceil \log |A| + 0.5 \rceil \geq 4$) or hashtags with a small number of adopters ($\lceil \log |A| + 0.5 \rceil \leq 1$) when all other baseline models fail to correctly classify any instances.
Basic network features are weak for viral memes, but good enough for dominant classes. Timing-based features work better for estimating future usage while distance-based features are more helpful for predicting the number of adopters. Community-based features yield the best results in general, particularly when detecting the classes of very popular memes. By combining all the features together, $P_n$ provides the best overall results. 
The network-based approach outperforms all baselines in detecting rare events---extremely popular and extremely unpopular hashtags.

\section{Conclusion}

In summary, we investigated the problem of predicting the future popularity of a
meme with three intuitive classes of features.
First, the positions of early adopters in the network provide information on 
the size of potential audience groups, which may affect the future popularity.
Second, community diversity is a good predictor of virality, consistently with 
prior findings that viral memes are less affected by community structure~\cite{weng2013viral}. 
Finally, the early growth rate of a meme usage  can be extrapolated 
to predict its future popularity, although the predictive power is not as strong as that of other features.

We have designed prediction features based on these intuitions and analyses, and tested them with machine learning techniques. The evaluation was executed against two simple baselines, as well as three more sophisticated regression models using early popularity (LN and ML models) or social influence of early adopters (social influence model).
The LN model has been shown to be a powerful predictor for inferring the future popularity of a single item, such as a tweet or a YouTube video, but does not perform well in predicting the popularity of hashtags. 
The ML model provides better results than the LN baseline by incorporating early popularity growth patterns.
The social influence model is able to achieve better performance than the LN model with knowledge of network structure and topological location of each early adopter.
However, none of the three regression models is capable of capturing the most popular memes nor the most unpopular ones.
Our prediction model outperforms all baselines in most cases, especially when predicting memes in the crucial minority classes. The performance is robust across different community detection methods.

Community-based features perform the best among the three classes. Predicting the number of meme adopters is a more difficult task, but our network-based approach outperforms other baseline models, especially in predicting memes with few adopters. The performance increases with longer observation windows.

The influence model and features of basic network topology, distance, and community structure require knowledge about the network and the positions of early adopters, while the LN model, ML model, and timing features need the timestamps of early messages containing the meme. 
Depending on what type of information is  available, one might choose different approaches.

The ability to predict whether a meme can go viral by just observing a few early messages provides us with many potential applications in social media analytics, marketing, and advertisement. This study offers not only novel, powerful features but also the first comprehensive analysis comparing multiple approaches for early prediction of viral memes.

\noindent \textbf{Acknowledgement} 
The work was supported in part by the James S. McDonnell Foundation and by NSF Grants No. 1101743 and 0910812.

\noindent \textbf{Conflict of Interest}
Y.-Y.A. declares competing financial interests: he co-founded a social media analytics company ``Janys Analytics'' in 2011 and is currently one of the major shareholders. The other authors declare no competing financial interests.


\bibliographystyle{aaai}
\bibliography{refs-final}

\end{document}